\documentclass[twocolumn,trackchanges]{aastex631}
\shorttitle{Broadband radio study of the NPS}
\shortauthors{Iwashita et al.}

\begin{document}


\title{Broadband radio study of the North Polar Spur:\\
Origin of the spectral turnover with insights into the X-ray and Gamma-ray spectra}

\author{Ryoji Iwashita}
\author[0000-0003-2819-6415]{Jun Kataoka}
\affiliation{School of Advanced Science and Engineering, Waseda University, 3-4-1 Okubo, Shinjuku-ku, Tokyo, Japan}

\author[0000-0002-4268-6499]{Yoshiaki Sofue}
\affiliation{Institute of Astronomy, University of Tokyo, 2-21-2 Osawa, Mitaka-shi, Tokyo 181-0015, Japan}




\begin{abstract}
The North Polar Spur (NPS) is a giant structure that is clearly visible in both radio and X-ray all-sky maps. We analyzed broadband radio observations covering a range between 22 MHz and 70 GHz to systematically analyze the thermal/non-thermal emissions associated with the NPS. We demonstrate that the radio emission of the NPS comprises synchrotron, free-free, and dust emission; however, synchrotron emissions dominate over other emissions, especially at high galactic-latitudes. Moreover, the synchrotron spectra exhibit a power-law behavior with $N(\gamma)\propto\gamma^{-s}$ ($s\simeq1.8-2.4$) up to a few GHz moderated by a turnover at $\nu_{\rm brk} \simeq 1$ GHz, above which the spectral index $s$ decrease by one. Assuming that the turnover is due to the electrons cooled by synchrotron radiation before escaping (or advecting) from the emission region, the magnetic field strength can be estimated to be $B\sim 8 \rm\mu G$ if the NPS is a distant structure that is near the Galactic Center (GC). However, an unreasonably strong $B\sim 114\rm\mu G$ is required if the NPS is near the local supernova remnant (SNR). The corresponding non-thermal energy stored in the NPS is $E_{\rm n/th}\simeq 4.4\times 10^{55}$ erg in the GC scenario, whereas $E_{\rm n/th}\simeq 4.1\times 10^{52}$ erg is difficult to explain with a single local SNR. We also estimated the gamma-ray emission associated with the NPS through inverse Comptonization of the cosmic microwave background (CMB), which peaks at 100 - 1000 keV with a flux of $\nu F_{\nu}\sim 10^{-9}$ $\rm erg\,cm^{-2}s^{-1}sr^{-1}$ in the GC model, and may be a good candidate for detection by future X-ray/gamma-ray observatories.

\end{abstract}




\section{Introduction} \label{sec:intro}

\begin{table*}
    \caption{Radio all-sky data used in this study}
    \label{table1}
    \centering
    \begin{tabular}{cclcll}
        \hline
        Product & Release & Observatory & Projection & Resolution(HEALPix) & Data Type\\
        \hline
        DRAO 22 MHz Map & 2019 & DRAO & $\ang{1.2} \times \ang{1.5}$ & nested,res 8 (Nside=256) & HEALPix\\
        All-sky 150 MHz Map & 2019 & Parkes,AUS & $\ang{5}$ & nested,res 8 (Nside=256) & HEALPix\\
        Haslam 408 MHz & 2014 & GER,AUS,ENG & $\ang{;56}$ & ring,res 9 (Nside=512) & Mollweide\\
        Dwingeloo 820 MHz Map & 2019 & Dwingeloo,NLD & $\ang{1.2}$ & nested,res 8 (Nside=256) & HEALPix\\
        1.4 GHz Continuum Map& 2001 & Stockert,Villa-Elisa & $\ang{;35.4}$ & nested,res 8 (Nside=256) & Mollweide\\
        2.3 GHz Continuum Map& 2019 & Hartebeesthoek,ZAF & $\ang{;20}$& ring,res 8 (Nside=256) & Mollweide\\
        23 GHz & - & WMAP & - & - & Mollweide \\
        30 GHz Full Channel Map & 2015 & Planck & - & nested,res 10 (Nside=1024) & HEALPix\\
        44 GHz Full Channel Map & 2015 & Planck & - & nested,res 10 (Nside=1024) & HEALPix\\
        70 GHz Full Channel Map & 2015 & Planck & - & nested,res 10 (Nside=1024) & HEALPix\\
        
        \hline
    \end{tabular}
    \textbf{References.} 22MHz: \cite{1999A&AS..137....7R}, 150MHz: \cite{1970AuJPA..16....1L}, 408MHz: \cite{2015MNRAS.451.4311R}, 820MHz: \cite{2012A&A...543A.103P}, 1420MHz: \cite{2012A&A...543A.103P}, 2300MHz: \cite{1998MNRAS.297..977J}, 23GHz: \cite{2011ApJS..192...15G}, Planck: \cite{2014A&A...571A..11P}

\end{table*}
Galactic all-sky survey observations have identified numerous giant structures across multiple wavelengths. In the microwave bands, the nature of the diffuse Galactic emission in the WMAP temperature anisotropy data was investigated, and ``Haze'' component was observed by \cite{2004ApJ...614..186F} or \cite{2012ApJ...750...17D}. Similar Haze structures have been observed in all-sky surveys conducted by Planck satellites \citep{2013A&A...554A.139P}, and they have been confirmed to extend from the Galactic Center (GC) to within the range of $|b| \sim \ang{35} - \ang{50} $and $|l|\sim \ang{15}-\ang{20}$ \citep{2008ApJ...680.1222D}. It is also suggested that the emission from the GC is hard-spectrum synchrotron radiation.
In the gamma-ray bands, the Fermi Gamma-ray Space Telescope discovered ``Fermi bubbles'', a giant bubble structure extending from the center of the galaxy toward the north and south \citep{2010ApJ...724.1044S}. This structure has sharp edges of radiation that extend approximately $\ang{50}$ above and below the GC, with a longitudinal width of $\sim\ang{40}$, exhibiting a bipolar symmetric structure. In the X-ray band, similar bubble structures were discovered by ROSAT and eROSITA. According to \cite{2020Natur.588..227P}, soft X-ray bubbles that extend approximately $14\,\rm kpc$ above and below the GC are not remnants of a local supernova, but a galactic-scale giant structure closely related to the features observed in gamma-ray bubbles. The authors estimate the energy of the X-ray bubbles to be $\sim 10^{56} \,\rm erg$, which is sufficient to perturb the structure, energy content, and chemical enrichment of the circumgalactic medium of the Milky Way. A common origin has been proposed for the two phenomena, the WMAP-Planck Haze and high-energy bubbles, and the spatial dimensions and locations of the haze in the microwave and Fermi bubbles are indeed compatible within the limits of the experimental error \citep{2018arXiv181205228R}.\\
\indent The  North Polar Spur (NPS)/Loop I is one of the most characteristic structures in Galactic all-sky maps and observed in both the radio and X-ray bands. Loop I is the largest northward emission that spans $\sim\ang{120}$ in the sky, and the brightest part of these regions is called the North Polar Spur (NPS). Although half a century has passed since its discovery, two competing ideas have been actively debated to postulate its origin. One of these is a local bubble near the solar system ($100 \sim 200 \rm \, pc$) \citep{1971A&A....14..252B}. According to this claim, the NPS/Loop I is attributed to supernova activity from the Sco-Cen OB association \citep{1996rftu.proc..249E,2018A&A...619A.120K}, and several authors have concluded that it is a collection of gas and dust expanded by a supernova explosion and stellar wind. In addition, the spatial nonuniformity of NPS/Loop I is another factor that supports this theory. Another idea is the remnant of active galactic nucleus (AGN) and/or starburst outflow from the GC over 10 Myr ago \citep{1977A&A....60..327S}. The recent discovery of a series of structures from the GC, such as the Fermi bubbles, popularized this theory. It has been suggested that NPS/Loop I is located along the edges of these galactic structures, indicating that bubble structures and NPS/Loop I have the same origin from the galactic explosions.

In the X-ray observation by Suzaku satellite, emission from NPS is well reproduced by the three-component thermal radiation: (1) Local Bubble and solar wind charge exchange (SWCX),  (2) thermal emission and Galactic Halo(GH), and (3) cosmic X-ray background (CXB) \citep{2013ApJ...779...57K,2018ApJ...862...88A}.
The NPS was represented by a thin thermal emission of $kT = 0.3$ keV during the ionization equilibrium process. This is above the temperature of a typical galactic halo ($kT = 0.2$ keV) and can be interpreted as a shock-heated GH gas. These authors concluded that the results suggest past activity in and around the GC.
In contrast, in radio observation, various analyses have been conducted since the discovery of NPS. \cite{1972A&A....21...61S} analyzed the linear polarization at 1415 MHz and estimated the distance to the NPS to be 50-100 pc based on the coincidence of the polarization directions of light and radio waves. 
\cite{2015ApJ...811...40S} created a Faraday rotation measure (RM) map of the NPS using data sets from 1280-1750 MHz, which indicated that a part of the NPS is a local structure of several hundred parsecs.
In \cite{2021ApJ...908...14K}, the relationship between radio and X-ray emissions of the NPS was discussed. The X-ray emissions are closer to the GC by $\sim \ang{5}$ compared with the corresponding radio emission and the radio and X-ray offsets in the NPS are attributed to the shock compression and heating of the halo gas during past galactic explosions. 

Although many NPS analyses have been performed at single or a few wavelengths (the literature above) in the radio band, but broadband analysis over multiple wavelengths has not yet been performed. In this study, we aim to clarify the NPS emission mechanism and obtain a closer insight into its origin by combining NPS radio data over multiple wavelengths and analyzing the corresponding spectra.

\section{Radio data}\label{sec2}
\subsection{Data Processing}\label{data}
The 6 all-sky maps from 22 MHz to 2.3 GHz were taken on the ground (Table\,\ref{table1}) while data at 23 GHz were obtained with the WMAP satellite. The $\rm 30\,GHz\cdot 44\,GHz\cdot 70\,GHz$ data by the Planck satellite was downloaded from NASA (\url{https://lambda.gsfc.nasa.gov}), while the Planck data was obtained from NASA/IPAC (\url{https://irsa.ipac.caltech.edu/Missions/planck.html}). The 23 GHz map was already separated into synchrotron and free-free radiation using the Maximum Entropy Method (MEM) analysis in \cite{2011ApJS..192...15G}. Table\,\ref{table1} lists the data used in this study. 

The unit for each all-sky data point is the brightness temperature $T_{b} = I_\nu \times c^{2}/2k_{B}\nu^2$, where $I_{\nu}$ and $k_{B}$ denote the brightness and Boltzmann constant, respectively. The brightness temperature is a physical quantity that describes the radiation intensity of an astronomical object. As shown in Table\,\ref{table1}, the all-sky map is displayed differently depending on the frequency, such as Mollweide and HEALPix. Therefore, in this study, the all-sky maps were converted to a Mercator diagram (Appendix: Figure \ref{app1}). 

The mean of the observed data within a pixel was taken as the measured value, and the standard deviation was the error at that pixel. In addition, the cosmic microwave background (CMB; $T_{\rm CMB} = 2.7$ K) was uniformly subtracted from the 22, 150, 408, 820, and 1420 MHz observed data. Because some of the brightness temperature data of Planck contained negative values that were consistent with zero within the uncertainty, all negative values were set to zero. When the lower limit of the error bar was negative, the upper limit was considered for analysis. 
 
\subsection{Observation Error}\label{Error}
The radio data for each frequency in the ground-based observations contained some errors, which, in addition to the errors introduced by data processing in Section \ref{data}, are mainly classified as scale and zero-level errors. Table \ref{table2} summarizes these errors in the observed data. However, \cite{1999A&AS..137....7R} does not mention zero-level or scale errors for the 22 MHz observations; instead, they used a value of $\pm5000$ K as a typical error. For the 150 MHz observations, \cite{2015ApJ...801..138P} calibrated the 150 MHz map by comparing the absolutely calibrated sky brightness measurements between 110 and 175 MHz calculated using the SARAS spectrometer, and \cite{2021ApJ...908..145M} proposed a calibration by comparing with absolutely calibrated measurements from an Experiment to Detect the Global EoR Signature (EDGES). 
We considered which of the three errors (standard deviation during data processing, scale error, and zero-level error) 
dominated the analyzed domain. Consequently, the standard deviation for each frequency was $\lesssim 10 \,\%$ in the NPS region and dominant at 22, 150, 408, and 820 MHz; however, at 1420 and 2300 MHz, the zero-level error was dominant. Therefore, standard deviations were used as errors for ground-based observations from 22 to 820MHz and for WMAP and Planck, whereas zero-level errors were used for 1420 and 2300 MHz. The original data were used for the two corrections at 150 MHz because exhibited a minor effect.
The WMAP separation of thermal/non-thermal emissions was performed using the maximum entropy method (MEM) of \cite{2003ApJS..148...97B}. This study states that the total observed galactic emission matched the MEM model by less than $1\%$, whereas the emission was separated into individual components with low accuracy. Therefore, we considered that these separations are sufficiently reliable.

\begin{table}
    \caption{Errors in Radio Data}
    \begin{center}
        \label{table2}
        \footnotesize{
        \begin{tabular}{llll}
            \hline
            Frequency & Scale Error [\%]& Zero-level Error & Supplement\\
            \hline
            22 MHz & - & - & $\pm 5000$ K \\
            150 MHz & $5\pm0.8$ & $21.4\pm8$ K & [1]\\
            150 MHz & $11.2\pm2.3$ & $0.7\pm6.0$ K & [2]\\
            408 MHz & $\pm10$ & $\pm3$ K & -\\
            820 MHz & $\pm6$ & $\pm0.6$ K & -\\
            1420 MHz & $\pm5$ & $\pm0.6$ K & -\\
            2300 MHz & $\pm5$ & $\pm0.5$ K & -\\
            \hline
        \end{tabular}
        }
    \end{center}
    \textbf{References.} 22MHz: \cite{1999A&AS..137....7R}, 150MHz: [1] \cite{2015ApJ...801..138P} [2] \cite{2021ApJ...908..145M}, 408MHz: \cite{2015MNRAS.451.4311R}, 820MHz: \cite{1972A&AS....5..263B}, 1420MHz: \cite{1982A&AS...48..219R}, 2300MHz: \cite{1998MNRAS.297..977J}
\end{table}

\section{Analysis and Results}
\subsection{Turnover Frequency}\label{freq}
To compare the differences in the spectral shape between high and low frequencies, we compared the power-law using frequency data at 22 MHz, 150 MHz, 408 MHz, 1420 MHz, 2300 MHz, and 23 GHz. All data were analyzed at a resolution of $\ang{5}$ in both the galactic-longitude and galactic-latitude to align with the 150 MHz resolution.

\begin{figure}
    \centering
    \includegraphics[width=1\columnwidth]{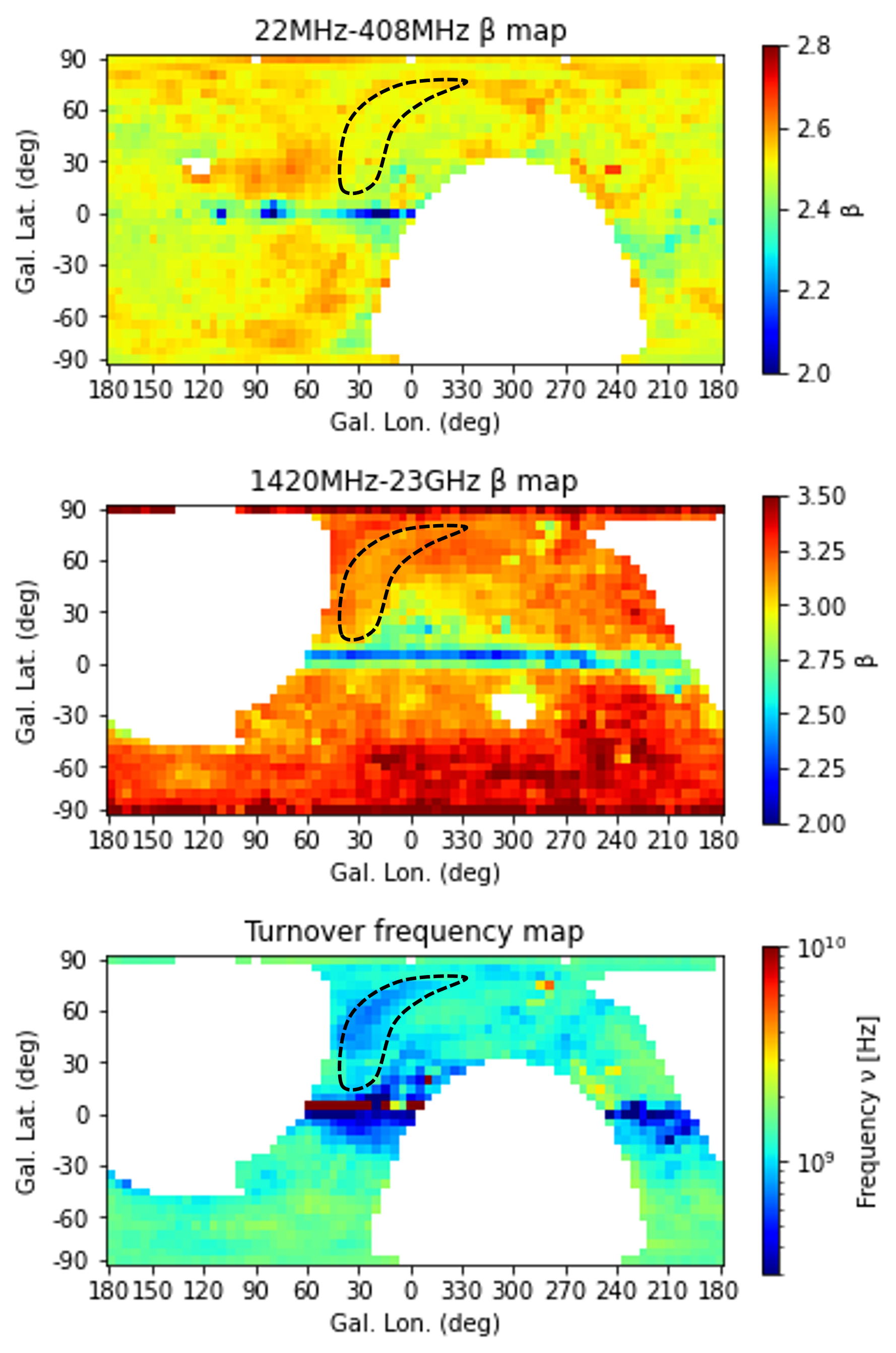}
    \caption{[top and middle] High and low frequency $\beta$ maps. [bottom] Turnover frequency map. The figures show the derivation of the power-law from 22 MHz to 408 MHz and from 1420 MHz to 23GHz, along with the turnover frequency. The black dotted line represents the NPS region.}
    \label{fig1}
\end{figure}

First, we obtained $\beta$ maps ($T_b = A\nu^{-\beta}$) using the least squares method with the brightness temperatures measured at different frequencies,

\begin{equation}
    \beta=-\frac{\sum_{i=1}^n \left( \ln\nu_{i}-\overline{\ln\nu}\right)\left(\ln T_{i}-\overline{\ln T}\right)}{\sum_{i=1}^n \left(\ln\nu_{i}-\overline{\ln\nu}\right)^2}
\end{equation}
where $\nu_{i}=$ 22 MHz, 150 MHz, and 408 MHz for a low-frequency map (Figure \ref{fig1}, top), whereas $\nu_{i}=$ 1420 MHz, 2300 MHz, and 23GHz for a high-frequency map (Figure \ref{fig1}, middle). As shown in Figure \ref{fig1}, the spectral power-law of the NPS region (white dotted line in Figure \ref{fig1}) was flatter than that in the other areas for both high and low frequencies. We obtained $\beta\simeq2.4$ to $2.7$ at low frequencies and $\beta\simeq2.7$ to $3.2$ at high frequencies in the NPS region, indicating that the power-law of the spectrum at high frequencies was steeper than at low frequencies.

Second, to understand the difference in the cutoff frequency of the radio spectrum in the all-sky, we obtained a turnover frequency map, as shown in the bottom of Figure \ref{fig1}. This figure plots the frequency at which the two power-law spectra obtained as shown at the top and middle of Figure \ref{fig1} intersect. For the two power-law functions obtained from 22 MHz to 408 MHz ($T_{1}=A_{1}\nu_{1}^{-\beta_{1}}$) and from 1420 MHz to 23 GHz ($T_{2}=A_{2}\nu_{2}^{-\beta_{2}}$), the turnover frequency at which the two functions intersect is 

\begin{equation}
\nu_{\rm{turnover}}=\left(\frac{A_{2}}{A_{1}}\right)^{\frac{1}{\beta_{1}-\beta_{2}}} [\rm Hz].
\end{equation}

As shown in the figure, the NPS spectral turnover was on the lower frequency ($<1$ GHz) than the other regions (see also \cite{mou2023cosmicray}). In reality, the turnover frequency in the NPS region for $l=\ang{30}-\ang{35}$, $b=\ang{55}-\ang{60}$ was 0.76 GHz, whereas it was 1.1 GHz for $l=\ang{40}-\ang{45}$ in the same galactic-latitudes around the NPS.

Next, in order to confirm the spectral turnover of the NPS region in more detail, the NPS spectra for each galactic-latitude are analyzed in the next section.

\subsection{Spectral Analysis}\label{spectrum}
\begin{figure*}
    \centering
    \includegraphics[width=1.8\columnwidth]{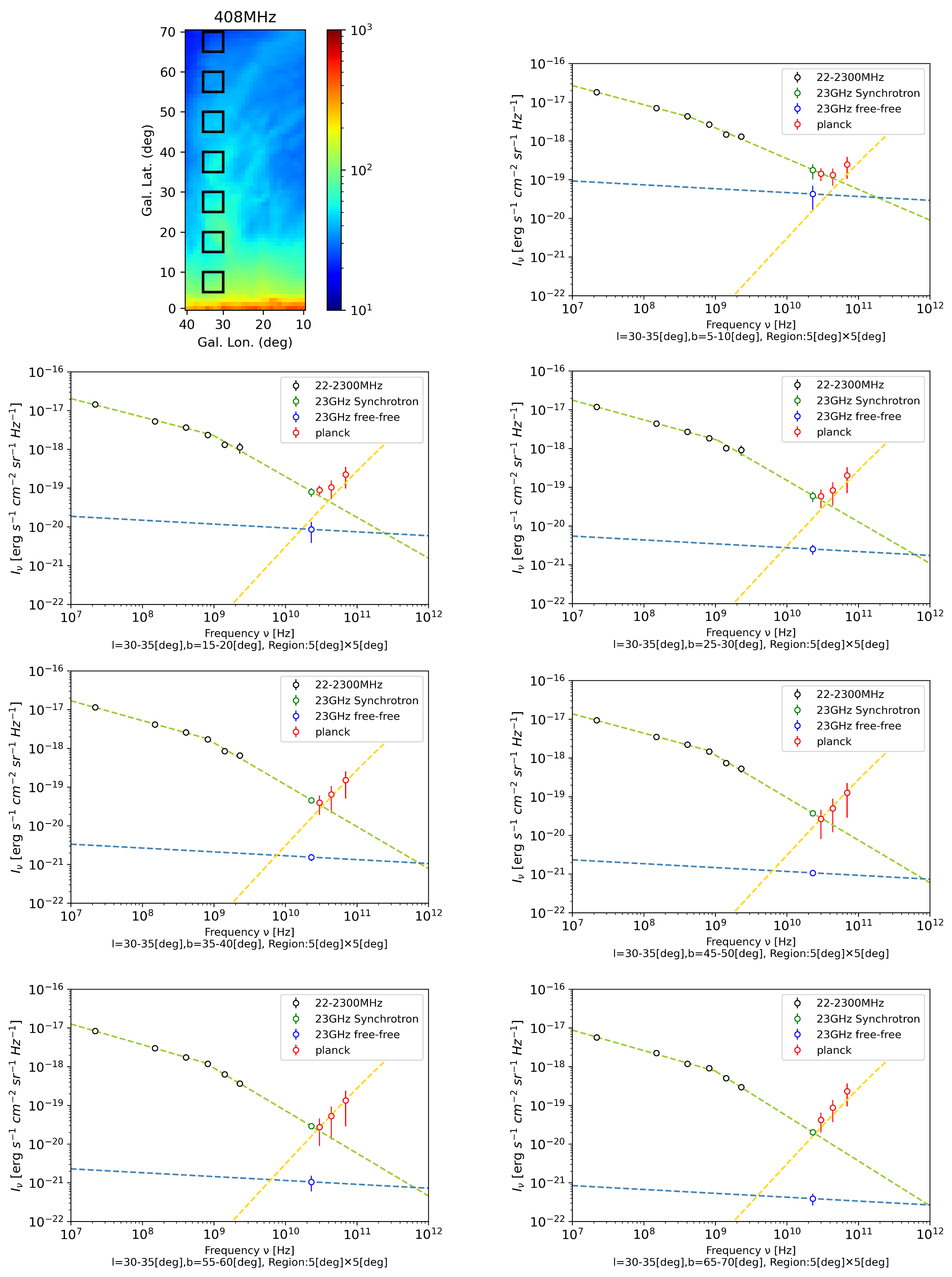}
    \caption{Spectra of the NPS in the radio band. The figure shows typical spectra for every $\ang{10}$ of galactic-latitude. The black, green, blue, and red (upper limit) points represent 22 - 2300 MHz, 23 GHz synchrotron radiation, 23 GHz free-free radiation, and Planck data (30 - 70 GHz), respectively. The green dashed line shows the synchrotron emission with the turnover obtained in section \ref{freq}. The blue dashed line shows the free-free emission with the $\alpha_{ff}=-0.1$ power-law. The yellow line is the dust emission assuming $\tau=5.0\times 10^{-6}$, and $T=20$ K.}
    \label{fig2}
\end{figure*}

We extracted the spectrum for each galactic-latitude of the NPS. To eliminate the effects of linear polarization and ensure that it was sufficiently larger than the beam width of all frequency maps, the spectral region was $\ang{5}$ in both the galactic-longitude and galactic-latitude. The galactic-longitude was fixed at $l=\ang{30}-\ang{35}$ where the NPS radiation is the brightest, and spectra were produced for regions varying in galactic-latitude by $\ang{10}$.
Figure\,\ref{fig2} presents the typical spectra for every $\ang{10}$ galactic-latitude in the NPS. The black, green, blue, and red points represent 22 - 2300 MHz, 23 GHz synchrotron radiation, 23 GHz free-free radiation, and Planck data (30 - 70 GHz), respectively. The NPS emissions were found to decrease at $\beta \simeq 2.4-2.7\,(T_b \propto \nu^{-\beta})$ power-law or $\alpha \simeq -0.4$ to $-0.7$ ($I_\nu \propto \nu^{\alpha}$, where $\alpha = 2-\beta$), up to a few GHz, regardless of the galactic-latitude. This result is consistent with the values in the literature $\beta \simeq 2.55-2.65$ \citep{2011A&A...525A.138G} or $\beta \simeq 2.3-3.0$ \citep{1988A&AS...74....7R}. Consequently, the electron spectrum indicates a power-law relationship with its index ($s$) of $N(\gamma)\propto \gamma^{-s}$ ($s\simeq 1.8-2.4$, where $s=1-2\alpha$) if synchrotron radiation dominated up to a few GHz. In addition, cut-offs were observed around $\nu_{\rm brk}\simeq1$ GHz, especially at high galactic-latitudes.

The synchrotron/free-free data at 23 GHz exhibited a stronger fraction of free-free radiation at low galactic-latitudes, and the contribution of free-free radiation decreased with increasing galactic-latitude. This indicated that synchrotron radiation dominated at high galactic-latitudes. Free-free radiation exhibited a flat power-law $\alpha \simeq -0.1$ in the optically thin region, whereas synchrotron radiation exhibited a significantly steeper power-law behavior with $\alpha > -1.0$. Therefore, synchrotron radiation clearly dominates at high galactic-latitudes and up to a few GHz.

We consider the 30 - 70 GHz Planck data where the radiation was consistent with the thermal radiation from dust with $\alpha \simeq 2.0$ power-law. In the optically thin limit, the Spectral Energy Distribution (SED) of the emission from a uniform population of grains is well described empirically by a modified blackbody $I_\nu = \tau_{\nu}B_{\nu}(T)$, where $\tau_{\nu}$ is the frequency-dependent dust optical depth and $B_{\nu}$ is the Planck function for dust at temperature $T$.

\section{Discussion}
\subsection{Radio emission from NPS}
\begin{figure}
    \centering
    \includegraphics[width=0.9\columnwidth]{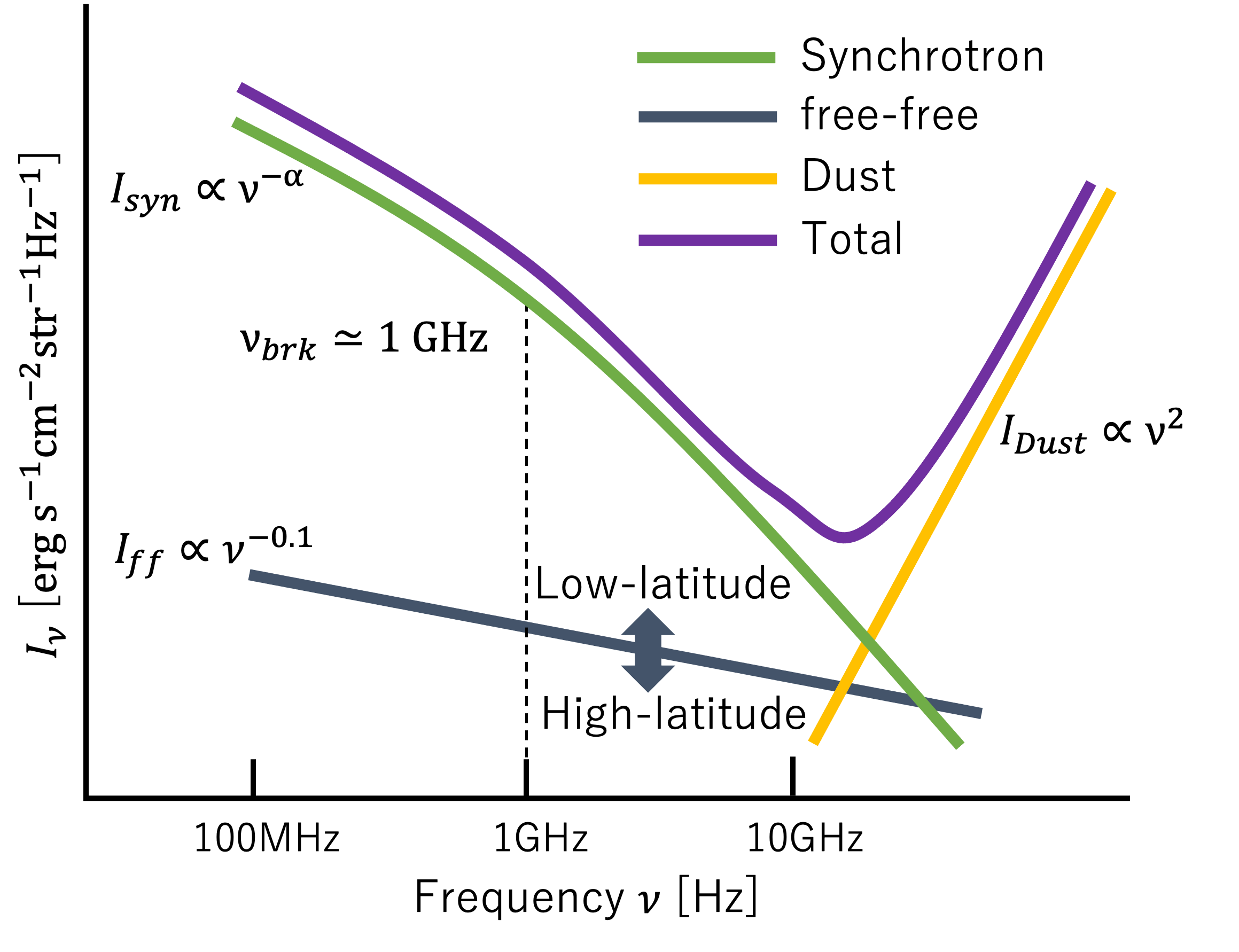}
    \caption{Characteristics of the NPS radio spectrum. (i)\,synchrotron radiation$ \,:\,$ dominant emission at low-frequency and high galactic-latitudes, with a cut-off around $\nu_{\rm brk}\simeq1$ GHz\, (ii)\,free-free radiation$\,:\,$ dominant emission at low galactic-latitudes\, (iii)\,Dust emission$\,:\,$ dominant emission at high-frequency.}
    \label{fig3}
\end{figure}

In Section\,\ref{spectrum}, we show that the radio emission of the NPS consists of (i) synchrotron radiation, (ii) free-free radiation, and (iii) dust emission. The characteristics of these types of radiation are described below. Synchrotron radiation decreases following a power-law behavior up to a few GHz and cut-off at $\nu_{\rm brk} \simeq 1$ GHz. Synchrotron radiation dominates up to a few GHz, whereas free-free radiation and dust radiation are almost negligible. Free-free radiation theoretically decreases at $\alpha_{\rm ff} \simeq -0.1$ power-law and can be observed from a few GHz. The contribution of free-free radiation increases, and the amount of radiation even exceeds that of synchrotron radiation, especially at low galactic-latitudes. This is clearly confirmed by the NPS spectra as shown in Figure \ref{fig2}. Dust radiation is dominant above several tens of GHz. In the optically thin limit, the SED of the emission from a uniform population of grains is well described empirically by a modified blackbody $I_\nu = \tau_{\nu}B_{\nu}(T)$.
Radio emissions of the NPS based on these characteristics are shown in Figure\,\ref{fig3}.\\

\subsection{Spectral turnover}\label{turnover}
The spectral turnover in the NPS can be discussed relative to synchrotron cooling. First, we assume the simplest Leaky-box model in which fresh, accelerated electrons are injected into an emission region with a magnetic field $B$, followed by the particle loss due to energy-independent advection and radiative energy loss. Approximating the transport equation for electrons passing through the shock by the Leaky-Box model, the time evolution of the electron energy distribution $N_{e}(\gamma)$ can be expressed as follows \citep{1996ApJ...463..555I},
\begin{equation}
    \frac{\partial N_{e} (\gamma)}{\partial t}=\frac{\partial}{\partial \gamma}\left(\frac{\gamma}{t_{\rm cool}} N_{e}(\gamma)\right) +Q_{0}(\gamma)-\frac{N_{e}(\gamma)}{t_{\rm adv}},
    \label{LeakyBox}
\end{equation}
where $t_{\rm cool}$ is the electron cooling time; $t_{\rm adv}$ is the advection time; and  $Q_{0}(\gamma)$ is the electron injection rate. The electron energy steady-state solution can be expressed approximately as a broken power-law of the form of Equation\,\ref{elecdist}
\begin{equation}
\begin{split}
    N_{e}&(\gamma > \gamma_{\rm min})\\
    &=N_{0}\gamma^{-s}\left(1+ \frac{\gamma}{\gamma_{\rm brk}} \right)^{-1} \rm exp\left(\frac{\gamma}{\gamma_{\rm max}}\right).
    \label{elecdist}
\end{split}
\end{equation}
Here, $s$, $N_{0}$, and $\gamma_{\rm min}$ (or $\gamma_{\rm max}$) denote the injection index, normalization constant, and minimum (or maximum) electron energy, respectively. Then, $\gamma_{\rm brk}$ corresponds to the energy at which the electron cooling time and advection time are balanced ($t_{\rm cool}\simeq t_{\rm adv}$), resulting in an electron distribution with a turnover whose power-law index $s$ steepens by one. If the spectral turnover $\gamma_{\rm brk}$ originates from synchrotron cooling, the magnetic field $B$ can be estimated by balancing the advection and cooling time scales of electrons. 

In a non-relativistic strong shock wave, the compression ratio is $v_{2}/v_{1}=1/4$, where $v_{1}$ and $v_{2}$ are the upstream and downstream velocities in the rest frame of the shock, respectively.
The electron advection time is obtained as follows:
\begin{equation}
\begin{split}
    t_{\rm adv}&\simeq \frac{R}{v_2}\\
    &\simeq 1.23 \times 10^{15}\left(\frac{R}{1 \rm\, kpc}\right)\left(\frac{v_{1}}{100 \rm\, km/s}\right)^{-1}\,[\rm sec],
\end{split}
\end{equation}
where $R$ denotes thickness of the radiation area. Using the literature value of $v_{1}\simeq 320$ km/s \citep{2013ApJ...779...57K}, in the GC model, and assuming that the distance to the radiation source was 8000 pc and that the NPS had a spherical structure with an outer diameter $R_{\rm out} \simeq 5$ kpc and an inner diameter $R_{\rm in}\simeq 3$ kpc ($R=R_{\rm out}-R_{\rm in}\simeq 2$ kpc) \citep{2015ApJ...807...77K}, the advection time of an electron was $t_{\rm adv}\simeq 7.7\times 10^{14}\,\rm sec \simeq 24.4$ Myr. In the SNR model, assuming that the distance to the radiation source was 150 pc, we obtained $R\simeq 2\,\rm kpc \times 150/8000 = 38$ pc, and the advection time of the electron was $t_{\rm adv}\simeq 1.5\times 10^{13}\,\rm sec \simeq 4.6\times10^{5}$ yr.

Meanwhile, the electron cooling time owing to synchrotron radiation is
\begin{equation}
\begin{split}
    t_{\rm cool}&\simeq \frac{E}{dE/dt}=\frac{\gamma m_{\rm e} c^{2}}{\frac{4}{3}\sigma_{\rm e}cU_{\rm B}\gamma^{2}}=\frac{6\pi m_{\rm e}c}{\sigma_{\rm e}B^{2}\gamma}\\
    &\simeq\frac{5.1\times10^{8}}{B^{2}\gamma}\,[\rm sec],
    \label{tcool}
\end{split}
\end{equation}
where $\sigma_{\rm e}$, $U_{B}$, and $\gamma$ denote the Thomson cross-section, magnetic field energy density, and Lorentz factor, respectively. A typical synchrotron frequency can be expressed as
\begin{equation}
    \nu_{\rm syn}=\frac{eB}{2\pi m_{\rm e}}\gamma^{2}\simeq 1.2\times 10^{6}B\gamma^{2}\,[\rm Hz],
    \label{nusync}
\end{equation}
where $B$ denotes the magnetic field in units of Gauss.
Then, by substituting Equation\,\ref{nusync} into Equation\,\ref{tcool}, we obtain
\begin{equation}
\begin{split}
    B\simeq&5.9\times \left(\frac{\nu_{\rm syn}}{1\,\rm GHz}\right)^{-1/3}\\
    &\times\left(\frac{v_{1}}{100\,\rm km/s}\right)^{2/3}\left(\frac{R}{1\,\rm kpc}\right)^{-2/3}\,[\rm \mu G].
    \label{cool_adv}
\end{split}
\end{equation}
Figure\,\ref{fig2} shows the presence of a cut-off at $\nu_{\rm brk}\simeq1$ GHz, which indicates that the electron cooling time and advection time are balanced ($t_{\rm cool}\simeq t_{\rm adv}$). Thus, by substituting the cut-off frequency $\nu_{\rm syn}\sim \nu_{\rm brk}\simeq1$ GHz and the electron advection time $t_{\rm adv}$ into Equation\,\ref{cool_adv}, the magnetic field of the NPS can be obtained, whose values are $8\rm \mu G$ and $114 \rm \mu G$ for the GC and SNR models, respectively. 
Note that the NPS radius varies by a few kiloparsecs in the literature for the GC model (cf. \cite{2021A&A...647A...1P}). When the shell thickness is doubled, $B\simeq5 \,\rm\mu G$ and $B\simeq72 \,\rm\mu G$ for the GC and SNR models, respectively, but we do not believe that these errors affect the discussion.

Last of all, we note that such spectral turnover may be owning to various other models. Firstly, if the low-energy electrons do not have sufficient time to establish a steady state relative to synchrotron cooling, but radiation loss time is sufficiently short for high-energy electrons, similar turnover could be observed around the GHz energy band.  
In such a case, we can no longer assume exquisite balance between $t_{\rm adv}$ and $t_{\rm cool}$; however, Equation \ref{tcool} and \ref{nusync} still hold, which leads to $B\simeq 68\times\nu_{\rm syn}^{1/3}\,[\rm GHz]\times t_{cool}^{-2/3}\,\rm[Myr]$. Therefore, if the NPS is a GC structure over $\simeq10$ Myr ago and the cooling time is comparable, $B\simeq 10\,\rm\mu G$ is naturally obtained. Similarly, we obtained $B\simeq 100\,\rm\mu G$ in the case of nearby SNR. Hence, we infer that our discussion will not be affected provided the cooling time is comparable to the dynamical time scale (or age) of the NPS.

Secondly, the bremsstrahlung energy losses may play a significant role in including spectral breaks, particularly in regions with high gas densities \citep{2010Natur.463...65C}. However, the bremsstrahlung energy loss timescale in the case of NPS is approximately three orders of magnitude longer than the synchrotron timescale, advection time, and age of the NPS because of its notably thin gas density of $n\simeq10^{-3} \,\rm cm^{-3}$ (e.g., \cite{2019MNRAS.485..924S}) compared to galactic center. Consequently, the effect of the bremsstrahlung energy losses can be regarded as negligible.

Finally, although there is certainly a steady state, bremsstrahlung collisions experienced by the electrons on the ISM gas are the dominant loss process for low-energy electrons. 
In fact, bremsstrahlung emission from non-thermal electrons is observed in a few SNRs, where a hard spectrum is possibly owing to the interaction of electrons with a thick interstellar medium and a gas density of approximately $n\simeq 10-100 \,\rm cm^{-3}$ (\cite{2002ApJ...571..866U}; \cite{2018ApJ...866L..26T}). Compared to this, the gas density in the NPS is sufficiently low, where $n\simeq 10^{-3} \,\rm cm^{-3}$ and $n\simeq 1 \,\rm cm^{-3}$ for the GC and SNR scenarios, respectively. Therefore, we infer that the contamination of the bremsstrahlung emission from non-thermal electrons cannot account for the spectral turnover as observed in GHz band.

\subsection{Spectral Energy Distribution}\label{SED}
\begin{table}
    \caption{Comparison of the fitting parameters of the SED}
    \begin{center}
    \footnotesize{
    \begin{tabular}{lll|l}
        \hline
        Parameter & NPS/GC&NPS/SNR&Bubbles\\
        \hline
         Distance\,[pc]&8000&150&8000\\
         $B$ [$\rm \mu G$]&8&114&10\\
         Region\,radius\,[cm]&$2.2\times10^{20}$&$4.0\times10^{18}$&$1.2\times10^{22}$\\
         Bulk\,Lorentz\,Factor&1&1&1\\
         $Q_{\rm 0}$\,[$\rm cm^{-3}\gamma^{-1}sr^{-1}$]&$4.0\times10^{-4}$&$5.2\times10^{-4}$&$1.7\times10^{-4}$\\
         $\gamma_{\rm min}$&1.0&1.0&1.0\\
         $\gamma_{\rm brk}$&$1.0\times10^{4}$&$2.7\times10^{3}$&$1.0\times10^{6}$\\
         $\gamma_{\rm max}$&$2.5\times10^{4}$&$6.0\times10^{3}$&$1.0\times10^{7}$\\
         Index $s$&1.9&1.9&2.2\\
         Brightness index $\beta$&2.45&2.45&2.6\\
         Intensity index $\alpha$&-0.45&-0.45&-0.6\\
         \hline
        $U_{e}$ [erg $\rm cm^{-3}$]&$1.2\times10^{-12}$&$1.2\times10^{-12}$&$1.8\times10^{-13}$\\
        $U_{B}$ [erg $\rm cm^{-3}$]&$2.5\times10^{-12}$&$5.2\times10^{-10}$&$4.0\times10^{-12}$\\
        $P_{\rm n/th}$ [erg $\rm cm^{-3}$]&$1.2\times10^{-12}$&$1.7\times10^{-10}$&$1.4\times10^{-12}$\\
         \hline
    \end{tabular}
    }
    \end{center}
    \textbf{Notes.} For the electron energy distribution, we assumed a standard broken power-low form $N_{e}(\gamma > \gamma_{\rm min})=N_{0}\gamma^{-s}(1+\gamma/\gamma_{\rm brk})^{-1}\times  \rm exp(-\gamma/\gamma_{\rm max})$. $U_{e},U_{B}$: electron and magnetic field energy densities $U_{e}=\int d \gamma m_{e}c^{2}\gamma N_{e}(\gamma)$ and $U_{B}=B^{2}/8\pi$. $P_{\rm n/th}$: the non-thermal pressure $P_{\rm n/th}=(U_{e}+U_{B})/3$.
    \label{tab2}
\end{table}

\begin{figure*}
    \centering
    \includegraphics[width=2.0\columnwidth]{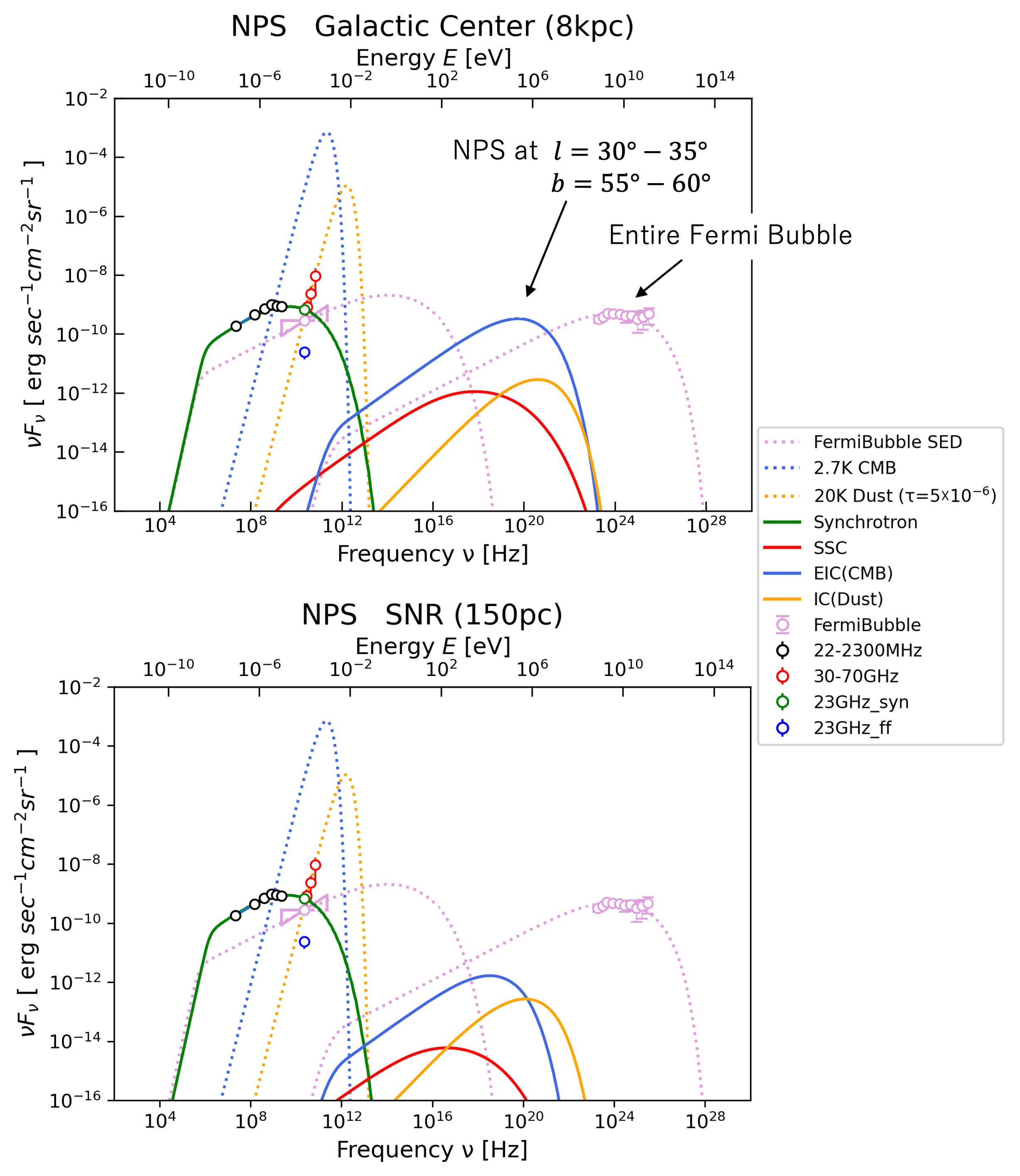}
    \caption{Spectral Energy Distribution of the NPS $(l,b)=(\ang{30}-\ang{35},\ang{55}-\ang{60})$ in the region of $\ang{5} \times \ang{5}$. Fitting was performed on data from 22 MHz – 23 GHz, which is pure synchrotron radiation, assuming distances to the NPS for each of the GC (8000 pc) and SNR (150 pc). The leptonic model was used for fitting \citep{1996ApJ...463..555I,1999ApJ...514..138K}. Green : synchrotron radiation, Blue : IC of the CMB, Red : Synchrotron-Self-Compton, Yellow : IC of the Dust, Purple : Fermi bubbles \citep{2013ApJ...779...57K}. }
    \label{fig4}
\end{figure*}

We performed a SED analysis for the high galactic-latitude region, where synchrotron radiation is dominant. We used the area\,$(l,b)=(\ang{30}-\ang{35},\ang{55}-\ang{60})$, where the spectral region is $\ang{5}$ in both the galactic-longitude and galactic-latitude. Fitting was performed on data from 22 MHz – 23 GHz, which is pure synchrotron radiation, assuming distances to the NPS for each GC (8000 pc) and SNR (150 pc). The leptonic model was used for the fitting \citep{1996ApJ...463..555I,1999ApJ...514..138K}, assuming the electron distribution as in Equation\,\ref{elecdist} and using the magnetic field obtained in Section\,\ref{turnover}.

The fitting results under these conditions are shown in Figure\,\ref{fig4}. The green, red, and blue lines represent synchrotron radiation, IC scattering when the CMB photons are knocked, and synchrotron-self-Compton (SSC) scattering. The yellow dotted line indicates the dust radiation, represented as $I_{\nu}=\tau_{\nu} B_{\nu}(T)$, whose optical thickness and temperature are $\tau_{\nu}\sim10^{-6}-10^{-5}$ and $T\sim20$ K, respectively, from the literature \citep{2014A&A...571A..11P}. Yellow is the IC when the dust radiation is knocked. In addition, the purple dotted lines and plots depict the SED of the Fermi bubbles fitted the one-zone leptonic model \citep{2013ApJ...779...57K}. The GeV data plots correspond to the emissions of the entire bubble structure, following \citep{2010ApJ...724.1044S}. The radio data plots correspond to the WMAP haze emissions averaged over $b = -\ang{20}$ to $-\ang{30}$ for $|l|<\ang{10}$. The bowtie centered at the 23 GHz K-band indicates the range of synchrotron spectral indices allowed for the WMAP haze, following \citep{2008ApJ...680.1222D}. The parameters used for the fitting and the corresponding results are listed in Table\,\ref{tab2}. Note that the best fit parameter, the hard electron spectrum $s\simeq1.9$, cannot be easily explained by the standard shock model. Although this is a very interesting topic, we simply adopted this value as it is beyond the scope of this study.

In the GC model, the IC of the CMB is dominated by high-energy radiation, with a peak in the 100–1000 keV range and flux of $10^{-9}$ erg\,$\rm s^{-1}cm^{-2}sr^{-1}$, whose brightness is almost equal to that of the high-energy band of the Fermi bubble. However, the IC of the CMB is dominant in the SNR model, similar to the GC model, but with a peak below 10 keV and flux of approximately $10^{-12}$ erg\,$\rm s^{-1}cm^{-2}sr^{-1}$. This value is several orders of magnitude lower than the detection limits of modern astronomical satellites. More accurate fitting will be possible if all-sky observations of gamma-rays progress in the future and NPS are discovered.

\subsection{NPS non-thermal Energy}\label{energy}
In Section\,\ref{SED}, owing to the distance to the two different NPS, we obtained the parameters shown in the Table\,\ref{tab2}, where $U_{e}$ and $U_{B}$ are electron and magnetic field energy densities.
\begin{align}
U_{e}&=\int d \gamma m_{e}c^{2}\gamma N_{e}(\gamma)\\
U_{B}&=B^{2}/8\pi
\end{align}

$P_{\rm n/th}$ is the non-thermal pressure, $P_{\rm n/th}=(U_{e}+U_{B})/3$. The total energy can be estimated by assuming NPS volume. In the GC model, the total volume of the NPS can be calculated as $V= 4\pi/3 \times (R_{\rm out}^{3} - R_{\rm in}^{3}) \sim 1.2 \times 10^{67} \,\rm cm^{3}$ using $R_{\rm out} \simeq5$ kpc and $R_{\rm in}\simeq3$ kpc \citep{2015ApJ...807...77K}. Thus, we estimated that the non-thermal energy of the NPS was $E_{\rm n/th}=V \times (U_{e}+U_{B})\simeq 4.3\times 10^{55}$ erg, which was consistent with the literature values $E_{\rm n/th} \sim 10^{55-56}$ erg \citep{1984PASJ...36..539S,2018Galax...6...27K,2021MNRAS.506.2170S}. 
It is notable that the NPS radius contains an error of several kiloparsecs in the GC model. When the shell thickness is doubled, some of the parameters listed in Table \ref{tab2} change to $Q=1.0\times 10^{-3} \,\rm cm^{-3}\gamma^{-1}sr^{-1}$, $\gamma_{brk}=1.2\times10^4$, $U_{e}=2.9\times 10^{-12} \,\rm erg\,cm^{-3}$, and $U_B=9.9\times 10^{-13} \,\rm erg\,cm^{-3}$, and the non-thermal energy is estimated to be $E_{n/th}\simeq1.3\times10^{56} \,\rm erg$. However, we do not believe that these errors to affect the discussion.

The thermal pressure of the NPS was estimated by \cite{2013ApJ...779...57K} using X-ray Suzaku observations. It should be noted that they are estimated using $\ang{8} < l < \ang{16},\, \ang{42} < b < \ang{48}$ observation data, and yielded $P_{\rm th}\simeq 2 \times 10^{-12} \,\rm erg\,cm^{-3}$. It was found to be in close agreement with the non-thermal energy determined for the first time from the radio spectrum of the NPS. Therefore, this can be interpreted as natural in the case of GC.
In contrast, in the SNR model, the volume of the NPS was $V \sim 7.9 \times 10^{61} \,\rm cm^{3}$, and the non-thermal energy was estimated to be $E_{\rm n/th}\simeq 4.1 \times 10^{52}$ erg. When the shell thickness is doubled, the parameters change to $Q=1.2\times 10^{-3} \,\rm cm^{-3}\gamma^{-1}sr^{-1}$, $\gamma_{brk}=3.4\times10^3$, $U_{e}=2.8\times 10^{-12} \,\rm erg\,cm^{-3}$, and $U_B=2.1\times 10^{-10} \,\rm erg\,cm^{-3}$, and the non-thermal energy is estimated to be $E_{n/th}\simeq4.8\times10^{52} \,\rm erg$. 
Compared to the GC, the distance to the NPS was approximately $\sim 1/50$ times larger, and the number density of the gas was $\sim 1/7$ times greater. Thus, the thermal pressure was estimated to be $P_{\rm th}\simeq 1.4 \times 10^{-11} \,\rm erg\, cm^{-3}$ which was approximately ten times larger than the non-thermal pressure. Therefore, the pressure balance was found to be disturbed in the SNR model. In addition, the typical energy of a supernova remnant is $E\sim 10^{51}$ erg, which is unnatural, at least for a single SNR. Therefore, if the NPS is a local structure, it could be a massive structure associated with a super bubble \citep{2018A&A...619A.120K}.

\section{Conclusions}
We analyzed broadband radio observations covering a range between 22 MHz and 70 GHz to provide a systematic analysis of the thermal/non-thermal emissions associated with the NPS. Spectral analysis showed that the radio emission of the NPS is composed of (1) synchrotron radiation, (2) free-free radiation, and (3) dust emission. Up to a few GHz, NPS emissions were found to decrease at $\beta \simeq 2.4-2.7\,(T_b \propto \nu^{-\beta})$, regardless of the galactic-latitude. Consequently, the electron spectrum exhibited a power-law relationship with its index (s) of $N(\gamma)\propto \gamma^{-s}(s\simeq 1.8-2.4)$. In addition, the cut-off in the radio spectrum was found to be approximately $\nu_{\rm brk}\simeq1$ GHz for the first time, which may indicate that the electron cooling time and advection time are balanced if the spectral turnover is derived from synchrotron cooling. Moreover, the magnetic field of the NPS was estimated as $B\simeq 8 \rm \mu G$ and $B\simeq 114 \rm \mu G$ in the GC and SNR models, respectively. 
In the SED analysis of the high galactic-latitude region of the NPS in the GC model, we estimated that the gamma-ray emission associated with the NPS, through IC of the CMB, peaked at approximately 100 - 1000 keV with a flux of $\sim 10^{-9}$ $\,\rm erg\,cm^{-2}s^{-1}sr^{-1}$. However, in the SNR model, IC of the CMB dominated, but with a peak below 10 keV and flux of approximately $10^{-12}$ $\,\rm erg\,cm^{-2}s^{-1}sr^{-1}$. Using the fitting results, the non-thermal energy of the NPS was calculated to be $E_{\rm n/th}\simeq 4.3\times10^{55}$ erg in the GC model, which was consistent with the literature values $E_{\rm n/th}\sim 10^{55-56}$ erg \citep{1984PASJ...36..539S,2018A&A...619A.120K}. Moreover, it was found that the thermal energy confirmed by \cite{2013ApJ...779...57K} in the X-ray was almost balanced by the newly obtained non-thermal energy in the radio bands. We obtained $E_{\rm n/th}\simeq4.1\times10^{52}$ erg in the SNR model, which indicates that if the NPS radiation originates from a local bubble, it could be a super bubble \citep{2018A&A...619A.120K}. 
Future deep surveys in MeV range, such as eASTROGAM \citep{2018JHEAp..19....1D} and DUAL \citep{2011SPIE.8145E..0EV}, may further clarify the origin of NPS .

\section*{Acknowledgements}
This work was supported by JST ERATO Grant Number JPMJER2102 and JSPS KAKENHI grant no. JP20K20923, Japan.
We thank an anonymous referee for his useful comments and suggestions to improve the manuscript.

\section*{Data Availability}
The radio all-sky survey data and Planck data are available from NASA LAMBDA \url{https://lambda.gsfc.nasa.gov} and NASA IPRC \url{https://irsa.ipac.caltech.edu/Missions/planck.html}, respectively.

\newpage
\appendix
\section{Radio all-sky maps}
\begin{figure}
\centering
\includegraphics[width=1.0\columnwidth]{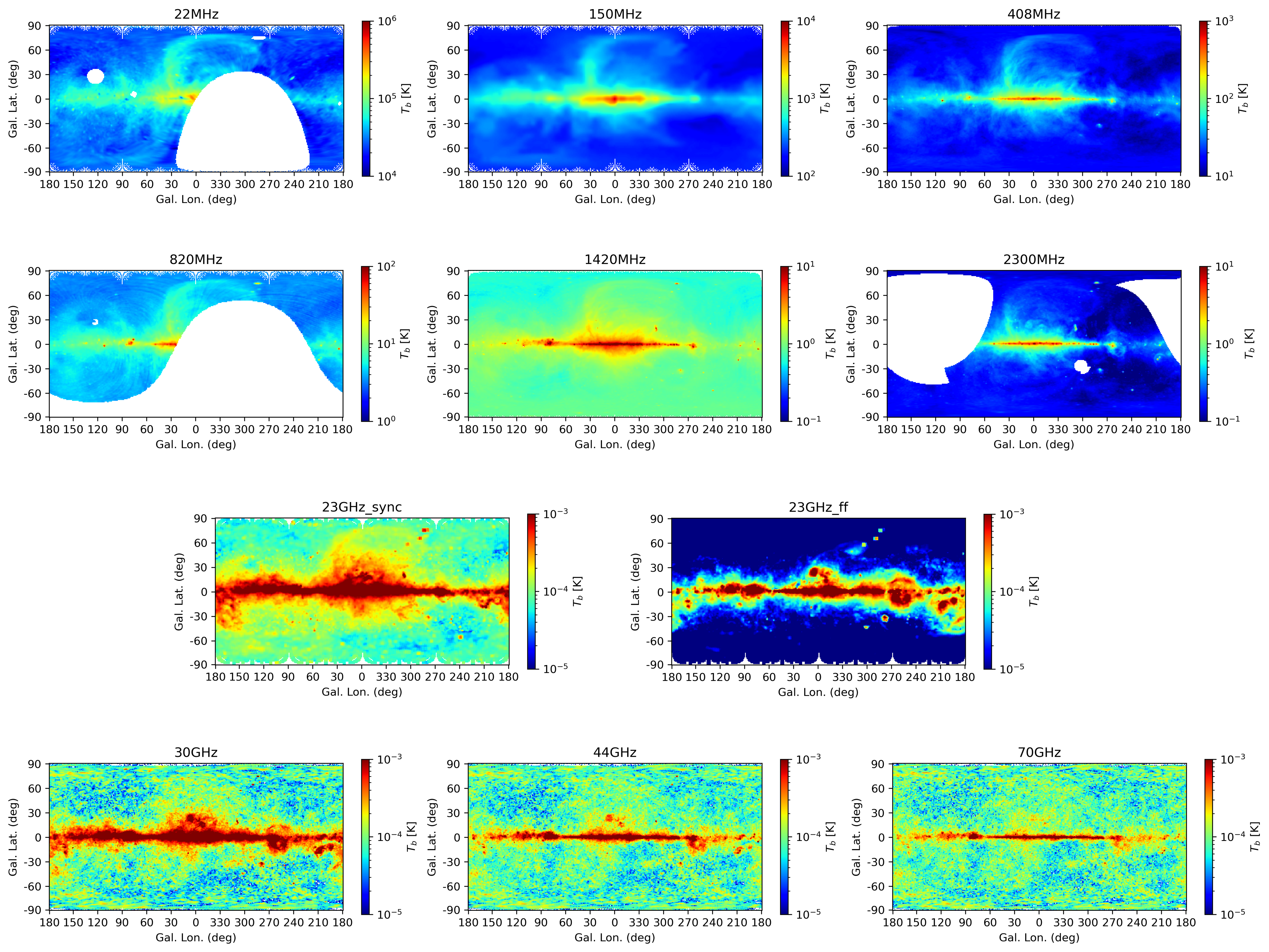}
    \caption{Radio all-sky maps from 22 MHz – 70 GHz. The all-sky maps were converted to the Mercator diagram, where resolution of all maps were converted to $(l,b)=(\ang{1} \times \ang{1})$ per pixel ($180 \times 360$ pixel).}
    \label{app1}
\end{figure}

The all-sky map used in this study is shown in Figure\,\ref{app1}. 
The method for creating an all-sky map is described in Section\,\ref{sec2}.

\bibliography{sample631}{}
\bibliographystyle{aasjournal}



\end{document}